# A Proposal for a FAIR Management of 3D Data in Cultural Heritage: The Aldrovandi Digital Twin Case

**Sebastian Barzaghi[1], Alice Bordignon[2], Bianca Gualandi[2,3†], Ivan Heibi[2], Arcangelo Massari[2], Arianna Moretti[2], Silvio Peroni[2], Giulia Renda[2]**

[1]Department of Cultural Heritage, Alma Mater Studiorum–Università di Bologna, via degli Ariani 1, Ravenna, Italy

[2]Department of Classical Philology and Italian Studies, Alma Mater Studiorum–Università di Bologna, Via Zamboni 32, Bologna, Italy

[3]Research Division (ARIC), Research Services and Project Coordination Unit, Alma Mater Studiorum–Università di Bologna, Via Zamboni 33, Bologna, Italy

## ABSTRACT

In this article we analyse 3D models of cultural heritage with the aim of answering three main questions: what processes can be put in place to create a FAIR-by-design digital twin of a temporary exhibition? What are the main challenges in applying FAIR principles to 3D data in cultural heritage studies and how are they different from other types of data (e.g. images) from a data management perspective?

We begin with a comprehensive literature review touching on: FAIR principles applied to cultural heritage data; representation models; both Object Provenance Information (OPI) and Metadata Record Provenance Information (MRPI), respectively meant as, on the one hand, the detailed history and origin of an object, and - on the other hand - the detailed history and origin of the metadata itself, which describes the primary object (whether physical or digital); 3D models as cultural heritage research data and their creation, selection, publication, archival and preservation.

We then describe the process of creating the Aldrovandi Digital Twin, by collecting, storing and modelling data about cultural heritage objects and processes. We detail the many steps from the acquisition of the Digital Cultural Heritage Objects (DCHO), through to the upload of the optimised

† Corresponding author: Bianca Gualandi (E-mail: bianca.gualandi4@unibo.it; ORCID: 0000-0001-8202-8493).

DCHO onto a web-based framework (ATON), with a focus on open technologies and standards for interoperability and preservation.

Using the *FAIR Principles for Heritage Library, Archive and Museum Collections* [1] as a framework, we look in detail at how the Digital Twin implements FAIR principles at the object and metadata level. We then describe the main challenges we encountered and we summarise what seem to be the peculiarities of 3D cultural heritage data and the possible directions for further research in this field.

## 1. INTRODUCTION

FAIR principles state that research data need to be Findable, Accessible, Interoperable and Reusable and consist in a list of 15 recommendations that are "related, but independent and separable" [2]. The authors remain rather generic and leave it to the different research communities to figure out how the principles can be put in practice in the respective disciplines and research workflows [3].

In 2020, the recommendations of the ALLEA E-Humanities Working Group defined data in the humanities as "all materials and assets scholars collect, generate and use during all stages of the research cycle" [3]. This definition seems to be increasingly accepted by humanities scholars [4], [5], [6], although some understandable resistance to the use of the term "data" remains [7], [8], [3].

The same report also highlights differences within the humanities, recognising that they themselves are diverse and that "data practices and demands vary significantly" [3]. All the same, many humanities researchers work with data made available by cultural heritage institutions which, while "not per se qualitatively different then data drawn from elsewhere", present some specific management challenges [1], [9]. The aim of this article is to shed light on some of these challenges, describing how they have been tackled in our use case, which concerns the creation of a *digital twin* of a temporary exhibition (now closed) dedicated to Ulisse Aldrovandi entitled "The Other Renaissance: Ulisse Aldrovandi and the wonders of the world"[1].

The creation of such a digital twin is part of the activities of the Project CHANGES ("Cultural Heritage Active Innovation For Next-Gen Sustainable Society")[2] and specifically its Spoke 4, which is a line of research of the project dedicated to investigating the use of virtual technologies for the promotion, preservation, exploitation and enhancement of cultural heritage in museums and art collections [10]. The original exhibition was held between December 2022 and May 2023 in the Poggi Palace Museum in Bologna, Italy, and consisted of a collection of more than 200 objects, largely never exhibited before, mostly belonging to the naturalist Ulisse Aldrovandi (1522-1605) and preserved by the University of Bologna [10].

[1] https://site.unibo.it/aldrovandi500/en/mostra-l-altro-rinascimento
[2] https://sites.google.com/uniroma1.it/changes/

This exhibition, made up of a large set of different small/medium objects, has provided an ideal experimental ground to define some approaches and methods relating to the acquisition, processing, optimisation, metadata inclusion and online publication of 3D assets, to be subsequently applied to Spoke 4's "core" case studies, which involve several "cultural institutions as representative of the different museum contexts in Italy" [10]. In this article we are going to focus on insights we have gained from the Aldrovandi Digital Twin pilot and in particular on the:

- Efforts made to produce cultural heritage data and metadata that are FAIR-by-design;
- Tracking of both the physical and digital Objects Provenance Information (OPI) and their Metadata Record Provenance Information (MRPI), respectively meant on the one hand as the detailed history and origin of an object, and - on the other hand - the detailed history and origin of the metadata itself, which describes the primary object (whether physical or digital);
- Peculiarities of 3D models, e.g. compared to 2D images, as cultural heritage data.

In doing so, this contribution seeks to answer the following research questions:

**RQ1:** What processes can be put in place to create a FAIR-by-design digital twin of a temporary exhibition?
**RQ2:** What are the main challenges in applying FAIR principles to 3D data in cultural heritage studies?
**RQ3:** From a data management viewpoint, how are 3D models "different from traditional digital images" [11]?

## 2. RELATED WORKS

### *2.1 FAIR principles and cultural heritage data*

Applying FAIR principles to humanities data, and cultural heritage data in particular, is not necessarily a natural fit [12]. It has been noted how, rather than being domain-independent, these principles "have been implicitly designed according to underlying assumptions about how knowledge creation operates and communicates" [9]. These assumptions–that there is a wide agreement on the definition of data, that scholarly data and metadata are digital by nature and that they are always created, and therefore owned, by researchers–certainly ring truer for traditionally data-driven disciplines than for the humanities.

It is now generally understood that FAIR principles should apply to all kinds of research objects [13], [14], [3], [12], but the effort to "translate" the principles to research workflows that are not traditionally data-driven is ongoing.

When considering cultural heritage data, two main obstacles to the application of FAIR principles have been highlighted: the "lack of explicit attention for long-term preservation", and the "excessive interwovenness of 'data' (or objects) and 'metadata'" [1]. In 2018 Koster and Woutersen-Windhouwer published a set of *FAIR Principles for Heritage Library, Archive and Museum Collections* that sought to emphasise the role of metadata by identifying three different levels of FAIRness: the object level (e.g.

books, artefacts, datasets), the object metadata level (e.g. title, creator) and the metadata records level (e.g. the body of metadata elements about the object) [1]. We will get back to this formulation of the principles in more detail in the discussion.

If we look at the wider ecosystem, other challenges to the widespread application of FAIR principles seem to be the lack of established and sustainable data curation workflows (e.g., cultural institutions mostly create siloed and static digital libraries) and the need for specific expertise within cultural institutions on how to digitise their holdings in a way that is useful to contemporary, often computational, research [15]. It is worth stressing that the availability of digitised cultural objects, even when paired with a download option, is not enough and that collections must be FAIR to be used as research data [1]. A possible solution to hasten innovation is to increase cooperation between galleries, libraries, archives, museums (GLAMs) on the one hand and researchers, data, information and computer scientists on the other [15]. As stated by the *Santa Barbara Statement on Collections as Data* [16] and, more recently, by the *Vancouver Statement on Collections as Data* [17], treating collections as data means encouraging the computational use of digitised and born digital collections but also committing to "respect the rights and needs of the communities who create content that constitute collections, those who are represented in collections, as well as the communities that use them" [17]. It entails many decisions relating to the selection, description, and access provision to be documented and shared and an alignment with emerging or established standards and infrastructures. Metadata are pivotal in developing trustworthy and long-lived collections, keeping in mind that it is an "ongoing process and does not necessarily conclude with a final version" [16].

Finally, the already cited DARIAH Position Paper *Cultural Heritage Data from a Humanities Research Perspective* reminds us that we need to "remain critical of equating virtual and physical manifestations and experiences of cultural heritage" [15]. Indeed, the recent *Cultural Heritage Image Sharing Recommendations Report* by the WorldFAIR Project states that:

> *[T]he GLAM [Galleries, Libraries, Archives and Museums] sector has a conceptual problem to overcome in the assumption that digital representations of images are mere surrogates for original objects. The digital files made available on image sharing platforms are unarguably primary research objects in and of themselves, and information about those objects is important to communicate* [18].

Therefore, the report recommends the adoption of a formal citation model for cultural heritage images [18]. Circling back to FAIR principles, and to cultural heritage objects in general, interesting work is currently being carried out within RDA, with the support of EOSC Future, to design an interoperable and FAIR-enabler citation model applicable to digital objects in the Cultural Heritage sector [19].

### 2.2 Metadata models and crosswalks

As Koster and Woutersen-Windhouwer [1] pointed out, the deep interconnection between the data or objects and the metadata describing them is a substantial trait of cultural heritage collections. In the digital domain, the presence of metadata enables the long-term preservation of cultural heritage collections, as well as the possibility of keeping track of both OPI and MRPI, while ensuring trustworthiness. Also, in this domain, it is crucial to provide insights into the evolution of data and metadata since often digital representations of artefacts evolve over time, either due to new discoveries or technological advancements in digitisation techniques.

As we have seen, working with data provided by cultural heritage institutions may present specific challenges. Part of the problem consists of the not-yet fully overcome conception that digital artefacts are just surrogates of the tangible original pieces. In addition, the majority of cultural heritage experts are not used to data-driven approaches and people who are experts on the data models and the information content of the metadata are often not proficient in the use of the technologies needed for metadata exchange, management, and interconnection.

Moreover, the varied nature of the cultural patrimony results in the proliferation of representation models and formats for the representation of digital artefacts. Indeed, because of the variety of materials preserved in archives, museums, and libraries, data and metadata are described with models that are inhomogeneous in purpose, degree of formalisation, and semantic richness, resulting in more complex mappings and exchanges. This condition highlights the need to make resources interoperable, which is the primary condition to exploit the potential joint use of variously formalised data [2].

As of today, projects aimed at enhancing cultural heritage in the digital ecosystem with the goal of producing FAIR data by design are still relatively few and highly complex. The adopted solutions are often ad hoc, and the absence of a generally shared formal workflow makes the process resource-intensive–as documented, for instance, in the outcomes of a project aiming at aligning Italian standards for describing works of arts and photographs into CIDOC-CRM [20]. In similar situations, clear guidelines for sustainable cultural digital artefacts generation and management would reduce costs, timelines, and risk of semantic loss, facilitating the usage of cultural heritage data as FAIR digital objects.

The close interconnection between digital cultural artefacts and their metadata, coupled with complex and heterogeneous forms of metadata creation, makes the issue of interconnection and information exchange between differently modelled knowledge systems a particularly thorny matter. Indeed, creating integrated knowledge systems in this domain often requires the use of data and metadata in formats other than those in which they originally were structured.

*Schema crosswalks* are particularly crucial to harmonise metadata formats and data models – indeed, they permit mapping between conceptualization systems that describe at least partially overlapping domains to identify points of contact and divergence, facilitating exchanges. In the past years, several tools have been developed to implement these crosswalks and include, for instance, the *RDF Mapping Language (RML)*[3] [21] and *SPARQL Anything*[4] [22].

[3] https://rml.io/
[4] https://sparql-anything.cc/

### 2.3 3D models as cultural heritage research data

Currently, there is a lack of literature delving into the FAIR management of 3D models as research data in the cultural heritage sector, perhaps because their uptake is low [18]. It has been argued that 2D digitisation should remain the priority of cultural heritage institutions in order to provide to the research community full-text access to documental heritage [15]. However, 3D models can bring to light different aspects of cultural objects and enable different types of analysis. A recent report suggests that "the purpose and value of 3D imaging [...] will probably be very different from traditional digital images" and may require different policies and technologies to make them accessible and preservable [11].

The Guidelines published by the Italian Ministry of Culture as part of the *Piano nazionale di digitalizzazione del patrimonio culturale* (*National cultural heritage digitisation plan*, in English), for example, include a short section on 3D digitisation, to be developed further in future versions of the document [23]. However, there is still a lack of shared standards for the creation, management, publication and exchange of digital heritage 3D assets, limiting access and use of published resources.

In their study, Bajena & Kuroczyński [24] highlight that metadata accompanying the models often does not allow a full assessment of the relevance of the model for further use scenarios. Additionally, a significant amount of digital assets is not made available to the scientific community as it is stored in private archives on personal computers [24].

There has been a growing recognition of the importance of scientifically publishing 3D models and the crucial role of metadata in standardising such a process. Blundell et al. [25] provide recommendations for metadata practices throughout the digital asset lifecycle, aiding in the organisation, verification, and access of 3D data. In their study, they highlight the need for different metadata requirements in various disciplines and industries that use 3D data and emphasise the need for further community-driven work to develop and refine a metadata standard that is truly cross-disciplinary and widely interoperable.

Fritsch [26] highlights the importance of well-documented and enriched 3D models of archaeological objects, including real-world coordinates and accurate colours, for the data to be used further. He also recommends making 3D data interactive and visible by embedding single 3D models in scientific articles. Indeed, a 3D web viewer is essential for publishing archaeological data, but additional elements beyond a compressed version of the 3D model must be considered; a practical solution can be providing both a "presentation model" and an "archiving model" of the 3D data [26]. Overall, precise publication guidelines, including file formats and uniform orientation, are needed to bridge the gap between 3D models, web viewer requirements, and compliance with FAIR principles.

Quantin et al. [27] presents a solution aligned with FAIR principles for long-term archiving combined with online publication of 3D research data in the humanities. The authors have developed a new metadata schema, aligned with standard vocabularies and mapped to the Europeana Data Model (EDM)[5], that allows for a more precise description of 3D models, greater openness to non-archaeological humanities fields, and better FAIR compliance. Although its limited applicability to completed projects and models, it provides a framework for organising and documenting 3D research data in the humanities, facilitating its long-term archiving and publication.

#### *2.3.1 3D data creation and selection*

Before discussing the methods used for 3D data preservation and publication, it is crucial to first address the problem of data selection. 3D models are the output of a process that uses several different data acquisition methods, such as scanning and image-based modelling. Depending on the acquisition technique, different data types are obtained. Cultural heritage artefacts' 3D capturing is today possible due to the accessibility of affordable, precise entry-level techniques that use the principles of Structure from Motion (SfM) or Structured Light Scanning (SLS). Light Detection and Ranging (LiDAR), computed tomography (CT), and industrial-grade high-resolution SLS are also being used more frequently, depending on the task at hand [28]. In all cases, content producers are generally encouraged to provide their data, both raw and final, as much as feasible [29].

The 3D content obtained from the survey generally goes through different phases: the processing of raw data will obtain a first full-sized processed model, then optimised to obtain a suitable and performing second version for web publication. When creating data, it is best practice to periodically save files to prevent data loss during processing. Thus, creators will have different versions of data or derivatives to store, which can be costly, especially for scanner raw data. According to a survey conducted in 2019 by Moore et al. [29] about 3D data management involving creators, repository managers, and creator-managers, 3D data storage and size were reported to be one of the biggest challenges to face. While maintaining two versions of a 3D model ought to be easy, offering raw data long-term archiving presents another difficulty [26]. The preservation of single raw scans and final point clouds for SLS is recommended, and the same goes for the source photos used for SfM (ideally in uncompressed TIFF format). In the digital archive or repository, the project, scans, photos, and registration metadata must also be linked to the data.

Despite storage costs[6], keeping both raw and processed data was reported to be the basic and essential step of data preservation in the scientific field by the Archaeology Data Service (ADS) and the Center of Digital Antiquity both regarding data obtained from laser scans [30] and close-range photogrammetry [31]. Although 3D technology has advanced significantly, it is likely it will improve further. Retaining the raw data may be helpful for future reprocessing, to obtain an even better final 3D model.

[5] https://pro.europeana.eu/page/edm-documentation

[6] [...] storage covers not only the size of the media on which data is stored and backed up, but also encompasses the ongoing periodic process of data refreshment (the movement of datasets to new hardware or software environments). While the cost of physical storage continues to decrease, that of refreshment and long-term curation—key factors in continuing to make data accessible and available—does not. In addition, in order to take advantage of technological advances and decreasing costs in certain areas, archives have to periodically upgrade systems or parts thereof." [29].

#### *2.3.2 3D data complexity and formats*

Depending on the environment of use and project goals, the selection of the best format to use for 3D data can change. A 3D file stores information in a machine-readable format that consists of binary or plain text data [32]. There are currently more than 140 3D formats, but this does not mean that all of them are or will remain popular [33].

Understanding the *complexity* of 3D data is extremely important and helps choose the appropriate formats for content preservation and dissemination. This choice depends especially on a) how many 3D features can be stored and which ones are lost during the conversion; and b) the interoperability of the format across different software.

Starting from (a), the main key features of a 3D file are *geometry*, *appearance*, *scene* [34], and *animation* [32].

Geometry consists of the surface data (polygons or faces), data points (vertices), and the capacity to modify the geometry after exporting. Storing geometry is the most basic function of any 3D file format. To fully describe a model's appearance, surface properties must be stored in addition to the 3D model geometry. A highly realistic 3D model can be produced by combining material properties, textures, and colour information.

The appearance is the texture that is mapped to a model's surface (including transparency, colour diffusion, and reflection), its environment, and lights (including colour and position). Data about the model's appearance can be encoded in different ways. One of the most used methods is texture mapping, the process through which a 3D model surface is rendered by projecting 2D images (texture) onto it. Texture mapping is supported by the majority of 3D file formats; however, depending on the format, the 2D image containing texture information could be kept in another file, stored commonly in PNG/JPG formats. Attributing a set of properties to each mesh face is another popular method of storing texture data. Colour, texture, and material type are examples of common attributes. Additional methods for affecting a model's appearance are the application of normal, bump, and transparency maps, which need a parameterization to be defined on the 3D model's surface. As with texture, these maps can be stored separately, depending on the format used. Finally, in the rendering process, "shaders" implement surface properties. Shaders are essentially collections of instructions that specify how each pixel or vertex should be rendered. In some cases, such as 3D printing, the appearance is not essential. However, in 3D artefacts for archeology documentation and visualisation, appearance is considered extremely relevant [26].

The scene includes information about light sources, cameras and other 3D models, if any. This is especially crucial if the model consists of multiple pieces that must be assembled in a specific order to create the scene. It should be noted that scene information is not supported by most 3D file formats. Frequently, this data is superfluous (except for some cases, such as video game design) and would unnecessarily increase the file size.

Similarly, animations are not compatible with all file formats. Nonetheless, several formats do contain animation data for use in applications where it is required, such as films' animated scenes, or video game design. Additional data must be stored for animation and interaction. These factors must be considered when evaluating archiving formats, metadata levels, and documentation requirements.

The second essential aspect affecting the choice of a 3D format is its interoperability across different software (b). To define a level of interoperability for each format, we have to define first if it is proprietary or neutral. Proprietary formats have been conceived to work on specific software, for example Blender's BLEND file and AutoCAD's DWG file. A neutral 3D format enables collaboration and flexibility, with different people working with different CAD software on the same file. STL, OBJ (ASCII variant), PLY, and 3MF, are examples of neutral formats. According to the All3DP platform, which is focused on 3D printing, the most common formats updated to 2023 are STL, OBJ, FBX, COLLADA, and 3DS [32]. Also, heritage communities are using glTF and PLY formats more frequently, according to a recent Sketchfab survey [33], [29]. However, 3D data type is now included in the Library of Congress's suggested format standards for 2020–2021[7]: among the most popular formats, OBJ, STL, and PLY are listed as acceptable formats for 3D object outputs from photogrammetry as of September 2020. Many of the mentioned formats, although not all of them, are program-neutral.

To facilitate the collaboration between different institutions and scholars involved in the field, and the long-term preservation of 3D data, the use of open, program-neutral file formats is recommended. Since neutral file formats are more interoperable and reusable, they also align better with FAIR principles. However, proprietary formats of raw data are generally preferred because, when converting raw data to non-proprietary formats, metadata embedded in the file is frequently lost [29]. In this context, choosing neutral formats that allow for structured and customised embedded metadata can facilitate the process (e.g. PLY), while preserving and making accessible the original proprietary files is also recommended. As of this publication, there is not a universally shared file format able to solve all the complexities involved, which makes interoperability especially challenging.

#### *2.3.3 3D model publication*

As mentioned above, web-based 3D viewers are essential for their larger-scale accessibility in terms of visualisation, interaction, and metadata enrichment. Yet, if the objective is to carry out a close examination, measurement, or detailed comparison it is still necessary to download the high-poly version of a model (if available) and open it with desktop programs like MeshLab[8] or CloudCompare[9] for scanned and photography-based models.

The most widely used commercial web-based 3D viewer available today is Sketchfab[10], which imposes size restrictions on the hosted models both on free and paid versions [29]. On the other hand, self-hosted viewers have been developed, such as Voyager[11] from Smithsonian, 3DHOP[12] from the Visual Computing Laboratory at the Institute of Information Science and Technology (CNR-ISTI), and the ATON[13] framework developed by CNR-ISPC [35]. A self-hosted viewer gives more control over the generated data and prevents size restrictions on the model. However, many models require an optimised or reduced version for distribution to meet hardware requirements or make loading times reasonable.

[7] https://www.loc.gov/preservation/resources/rfs/design3D.html
[8] https://www.meshlab.net/
[9] https://www.cloudcompare.org/
[10] https://sketchfab.com/

Champion and Rahaman [33] analysed the most popular available commercial and institutional platforms for downloading, trading, sharing, and hosting 3D models, defining some useful features in the scholarly field of 3D digital heritage. Despite the growing number of 3D models, there are still few online libraries and accessible platforms on which the GLAM sector can rely on. Moreover, the functions offered are still limited and under development, and often they fail to offer pertinent data and links to additional resources for usage and research. Most commercial model platforms, including TurboSquid[14], CGTrader[15], and ShareCG[16], generally lack both OPI and MRPI. Surprisingly, neither institutional nor commercial options typically have integrated 3D viewers: in most cases, a 2D image or video preview is provided before downloading any 3D content.

### *2.3.4 3D model archival and preservation*

Regarding the availability of infrastructures for the long-term preservation of 3D cultural heritage data, a search on Re3Data[17], an internationally recognised registry of research data repositories, shows a total of only 23 results. The search has been carried out by typing the string "3D" in the search bar–as there is no suitable filter in the "Content Type" list–and by selecting "Humanities and Social Sciences" from the available subject areas in order to filter out infrastructure dedicated exclusively to 3D data in the natural science, medicine, engineering and more.

Of the 23 repositories found on Re3data:

- Three have a specific focus on 3D data[18].
- Three are institutional repositories dedicated to different types of research data, including 3D[19].
- Three repositories focus on open data from specific cities or regions of the World[20] and seem to include 3D among many other types of data and only on the condition they are published under open licences.

[11] https://smithsonian.github.io/dpo-voyager/
[12] https://3dhop.net/
[13] https://osiris.itabc.cnr.it/aton/
[14] https://www.turbosquid.com/it/
[15] https://www.cgtrader.com/
[16] https://www.sharecg.com/
[17] https://www.re3data.org/
[18] They are Open Heritage 3D (https://openheritage3d.org/), MorphoSource (https://www.morphosource.org/) and the French National 3D Data Repository (https://3d.humanities.science/).
[19] They are linked to cultural heritage institutions in two cases, the Smithsonian (https://www.si.edu/openaccess) and the Bayerische Staatsbibliothek (https://www.digitale-sammlungen.de/de), and to Washington University (https://data.library.wustl.edu/) in the other.
[20] Coquitlam Open Data (https://data.coquitlam.ca/), Thunder Bay Open Data Portal (https://opendata.thunderbay.ca/), Open Data DK (https://www.opendata.dk/).

- Two have a disciplinary scope, specialising in geographical and archaeological data, including in 3D format[21].
- Two can be considered generalist repositories, without a specific disciplinary or institutional mission[22].
- Finally, nine repositories are linked to research projects and host the relative research data, including 3D. These all belong to the publisher Topoi Edition[23], which has its own entry on the list of results, bringing them to 23.

This short survey is of course partial and does not depict a complete picture of the current situation. On the one hand, these repositories do not necessarily host actual 3D models, they just include somewhere in their documentation the term "3D". However, the lack of a suitable content type that can be used to filter the results is probably a sign of how marginal 3D content still is. On the other hand, although Re3data is a reliable tool[24], the search does not capture all infrastructure capable of hosting 3D content. For example, both Europeana[25] and Zenodo[26] are not included in the list of results although they accept 3D model uploads.

## 3. PROPOSED APPROACH: THE ALDROVANDI DIGITAL TWIN PILOT

The original list of FAIR principles [2] has in time been customised to adapt to other research objects – one well known example being research software [36]. Indeed, a few years earlier, Koster and Woutersen-Windhouwer [1] had proposed their own version of the principles for library, archive and museum collections, summarised in Table 1 below.

In this article we discuss the application of this extended version of the FAIR principles to 3D cultural heritage data through the lens of the Aldrovandi Digital Twin introduced in Section 1. The next section describes the workflow for the creation of the digital twin with special attention to the creation, reuse and management of cultural heritage data and to the application of FAIR principles throughout. In it, we try to follow the chronological order as much as possible. In Section 5, we answer our research questions, highlight the challenges that emerged in the process and we suggest some directions for future work.

[21] They are ArcGIS Living Atlas of the World (https://livingatlas.arcgis.com/en/home/) and the Archaeological Map of the Czech Republic (https://digiarchiv.aiscr.cz).
[22] Open Science Resource Atlas (https://zasobynauki.pl/) and PAC-Archiving Platform CINES (currently unavailable).
[23] https://www.re3data.org/repository/r3d100012470
[24] For example, it has recently been used, alongside FAIRsharing (https://fairsharing.org/), in a study commissioned by the European Research Council Executive Agency on the readiness of research data repositories.
[25] https://www.europeana.eu/it
[26] https://zenodo.org/

**Table 1.** FAIR principles customised for library, archive and museum collections [1].

| | Objects | Metadata about the object (elementary level) | Metadata records |
|---|---|---|---|
| **Findable** | Objects (physical and digital) have a globally unique persistent identifier.<br>Objects are described with metadata. | Metadata specify the global persistent identifier (PID) of the object. This PID is used in all systems/databases that contain a description of the object in question.<br>Metadata about specific objects are available via one or more searchable online repositories, catalogues, online databases, etc. | Metadata records have their own global persistent identifier. |
| **Accessible** | Digital objects are permanently accessible by:<br>• Sustainable storage (hardware, storage medium)<br>• Open universal access protocols<br>• Version management<br>• Backups. | Metadata specify information about an object's availability, obtainability and/or access options. | Metadata records are machine readable.<br>Metadata records are accessible using open universal protocols. |
| **Interoperable** | Digital objects are stored in preferred or acceptable formats. | Metadata are available at least in one metadata schema appropriate for the specific type of object.<br>Metadata are available in various additional generic standard data formats for other contexts.<br>Metadata contain links/references to other objects/authority files, by using other global persistent identifiers. | Metadata records must be of sufficient quality. |
| **Reusable** | Digital objects have a date-timestamp.<br>Objects have a licence for reuse, which is also available in a machine readable form. | Metadata specify the object's rights holder.<br>Metadata contain licence information referring to the object.<br>Metadata specify the object's provenance. | Metadata specify the metadata record's provenance.<br>Metadata specify the entity responsible for the metadata record.<br>Metadata records have their own licence for reuse, which is also available in a machine-readable form. |

## 4. METHODOLOGY

### *4.1 Collecting and storing data about objects and processes*

The digitisation workflow started with the creation of two tabular datasets: one for storing bibliographic and catalogue data (*bibliographic data* from now on) about the physical *cultural heritage objects* (*CHOs* from now on) included in the temporary exhibition, and another for data concerning the acquisition and digitisation process (*process data* from now on). Process data were eventually instrumental to the creation of the metadata accompanying 3D models representing the *digital cultural heritage object* (*DCHO* from now on) created in the digitisation process.

The table structure was conceptualised, defining column names, expected cell data, and controlled values for specific columns (e.g. object type). To facilitate the process of data entry in terms of technical skills, timing, and collaborative needs, both tables were hosted on Google Spreadsheet. This decision was motivated by several reasons. First, its familiar and intuitive interface allowed team members with minimal data management experience and under severe time constraints to easily collect data without extensive training. Moreover, its support for real-time collaboration enabled multiple contributors to work on the data at the same time, thus streamlining the workflow and reducing potential delays. In addition, its built-in version control ensured that changes to the data were tracked securely over time. While not ideal in terms of data management, it provided an overall cost-effective solution that offered essential functionalities for the project while being accessible, extensible and reproducible for any other team dealing with digitisation endeavours. The two tables were populated in parallel, paying great attention to consistency. Official sources–such as the official exhibition catalogue, preliminary unstructured notes created by exhibition curators as drafts of the exhibit organisation, and other museum cataloguing records–were used to collect bibliographic data, while the collection of process data was performed during the acquisition and digitisation activities.

Process data column names were structured according to the various steps of the acquisition and digitisation process and included all the relevant attributes to track each step. In particular, as summarised in Figure 1, the *Acquisition phase* (step 1) aimed to capture CHOs and the raw material (*RAW* from now on) needed to create the related DCHOs. The information related to the acquisition phase included: the *unit* responsible for working with a CHO; the *people* responsible for carrying out its acquisition; the *technique* used to capture the raw material of the CHO; the *tools* used to perform the acquisition activity; and the *dates* when the acquisition process was initiated and completed.

The acquisition phase was then followed by a series of software activities (steps 2-7), i.e. the phases that involved various software tools and applications to process, transform, and publish the digital versions of the raw material acquired in the previous phase. Although the software activities can vary depending on the nature of the materials being digitised and the intended use of the digital files, our tracked work included the following phases in the digitisation process of all objects: *Processing phase* (step 2), to manipulate the RAW produced during the acquisition phase, using automatic algorithms (where the human operator sets only their input parameters), to obtain a first processed raw model (*RAWp* from now on); *Modelling phase* (step 3), where the human operator fixes, as part of his/her authorial activity and subjective interpretation, possible topological issues of RAWp, obtaining the DCHO; *Optimisation phase* (step 4), to simplify the DCHO for specific purposes or use cases, obtaining a optimised DCHO (*DCHOo* from now on); *Export phase* (step 5), to convert the RAWp, DCHO and DCHOo into a specific format; *Metadata creation phase* (step 6a), to create structured information of bibliographic and process data of the CHO, RAW, RAWp, DCHO, and DCHOo; *Provenance creation phase* (step 6b), focused on the creation of MRPI to track the agent responsible for bibliographic/process data creation, the time of data creation, and the primary source of the data; *Upload phase* (step 7), to the DCHOo from a local device or storage location to a Web-based framework (e.g. ATON).

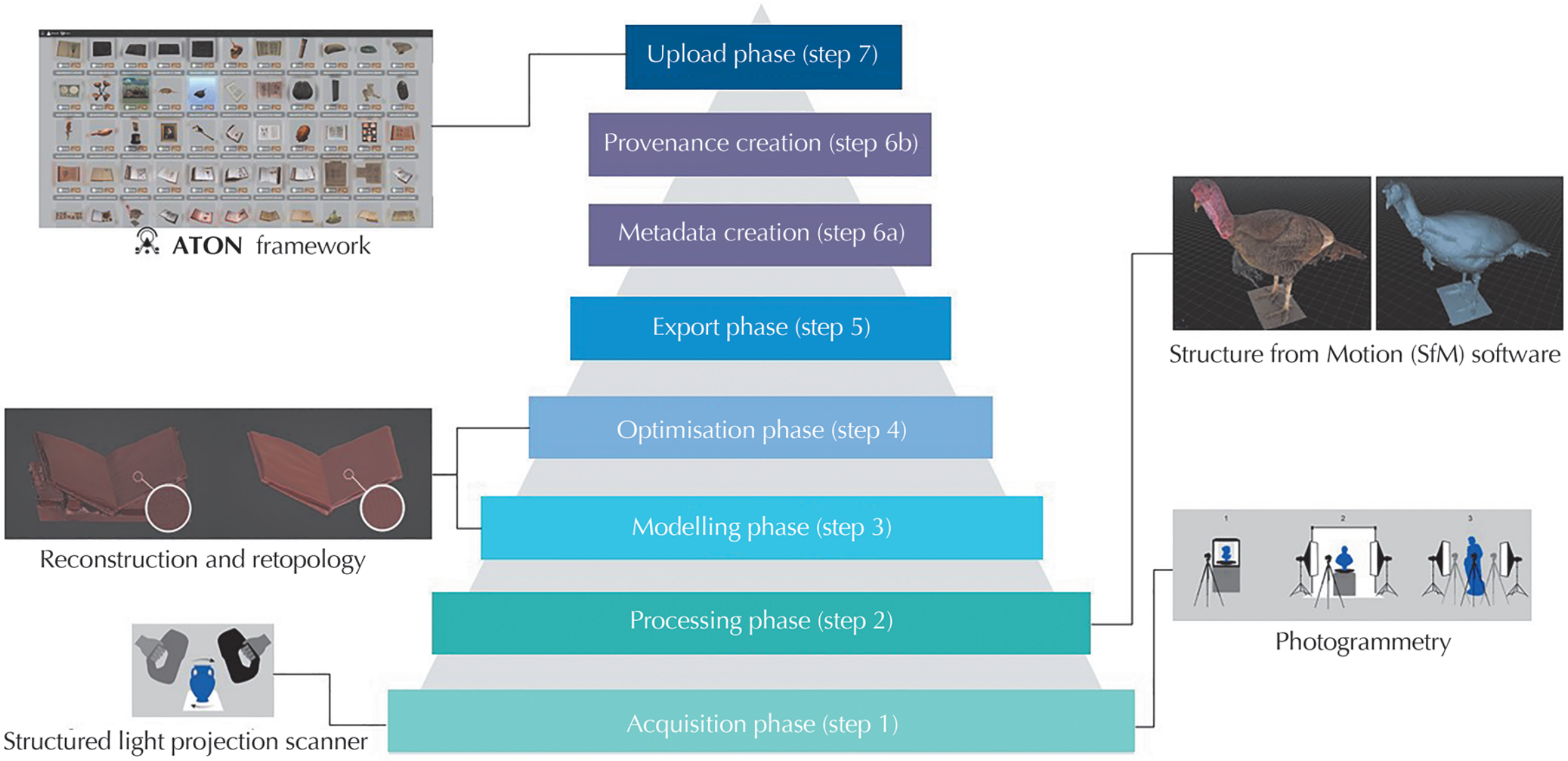

**Figure 1.** The acquisition and digitisation process (described in more detail below).

The RAW, RAWp, DCHO and DCHOo created and the related metadata were pivotal in creating an exhaustive record of the entire digitisation process, facilitating monitoring of each CHO/RAW/RAWp/DCHO/DCHOo status and evaluating the project's overall success. Additionally, they ensure the long-term preservation and accessibility of DCHOs and allow for in-depth analysis of the acquisition and digitisation dynamics, which can vary significantly in duration, depending on the project scope and the unique characteristics of each CHO.

### *4.2 Modelling data about CHOs, processes, provenance, and documentation*

As information was entered in both bibliographic and process data tables, a further effort was made to prepare them to be published as machine-actionable data [37] compliant with FAIR principles, adopting a formal, accessible, shared, and broadly applicable language for knowledge representation [2].

To exploit the bibliographic and process data tables, the information structured in tabular format had to be converted into a Resource Description Framework (RDF) format [38]. Moreover, to maximise data reuse and interlinking to other existing resources, certain data values were aligned with terms of existing knowledge graphs (such as WikiData[27]) and identifiers defined by authority lists (e.g. VIAF[28] and ULAN[29]). Thus, starting from the information used to populate the table cells, a OWL ontology called *Cultural Heritage Acquisition and Digitisation Application Profile* (CHAD-AP)[30] was developed for describing cultural heritage digitisation data and processes [39]. More precisely, CHAD-AP can be logically split into two separate abstract modules:

- an Object Module (OM) based on the CIDOC Conceptual Reference Model (CIDOC CRM)[31] [40] and designed from the bibliographic data to describe the CHO's characteristics and contextual information;
- a Process Module (PM) based on both CIDOC CRM and its extension CRM Digital (CRMdig)[32] [41] for describing the process data.

CHAD-AP was developed with the Simplified Agile Methodology for Ontology Development (SAMOD) [42], an iterative process for developing fully documented and tested ontologies, thus ensuring that both the model and the development process itself are accountable and reproducible. Moreover, the reuse of CIDOC-CRM as a foundational framework provided a well-founded and shared structure, largely adopted in cultural heritage projects, to standardise the semantics of both the CHO and the process for creating the related RAW, RAWp, DCHO and DCHOo, represent their nuances, and promote their interoperability with other existing models in the cultural heritage domain.

MRPI is systematically mapped via the OpenCitations Data Model (OCDM) [43]. In this framework, every time an entity is created or modified, a detailed record known as a "snapshot" is captured and preserved within a designated provenance graph. A snapshot acts as a historical marker, encapsulating the state of an entity at a specific point in time. Each snapshot is linked to its corresponding entity and records specific timestamps: the dates of their creation and when they become invalid, when applicable. The individuals responsible for the creation or modification of the entity's data are recorded as well, ensuring accountability and transparency, together with the source of information for tracing back to the primary sources of the data. Continuity in the historical evolution of each entity is maintained by connecting each snapshot to its previous version. In addition, we record change tracking activities of the entity in time, thereby facilitating the restoration of entities to specific snapshots by reversing operations from all subsequent updates.

The integration of the *Provenance creation* phase (step 6b) into the metadata creation process has significantly bolstered the FAIR management of the CHO, RAW, RAWp, DCHO and DCHOo and their bibliographic and process data. This robust mechanism for tracking the history and changes of all these entities ensures their authenticity and reliability, and therefore their reusability in various contexts, especially in research.

[27] https://www.wikidata.org/
[28] https://viaf.org/
[29] https://www.getty.edu/research/tools/vocabularies/ulan/
[30] https://w3id.org/dharc/ontology/chad-ap
[31] http://www.cidoc-crm.org/cidoc-crm/
[32] http://www.ics.forth.gr/isl/CRMdig/

Finally, we use MELODY, an online dashboarding system for creating web-ready data stories that leverage Linked Open Data [44], for providing human-readable documentation of all the metadata tracked in RDF and to enable the creation of narratives introducing the CHOs considered and the DCHOs created.

### *4.3 Open technologies and standard formats for interoperability and preservation*

Our aim has been to use open technologies and software at every stage of the process to maximise the workflow's transparency and re-adoption in different scenarios, both internally or externally to Spoke 4. While metadata did not present challenges from this point of view, working with 3D models did. A selection of open-source software was performed for all the steps of the process. However, proprietary software was needed for some specific tasks, where open-source software does not provide satisfying results.

To avoid being restricted to any proprietary software applications we used as many standard formats as possible for all the different types of research data we produced. These include 3D models (glTF, GLB, obj, mtl, png, jpg, tiff, e57), images (png, tiff, raw, jpg,), video (mov, mp4), and audio (mp3). These choices were recorded and in some cases guided by the project's Data Management Plan [45].

We pushed for the adoption of glTF for 3D formats, an open standard that is interoperable and aimed at interactive Web3D applications, to ensure high interoperability with current 3D platforms and services, as well as re-use and integration of licence data within the format [10], [46].

The final steps of the projects are still ongoing. As we write, the 3D models and accompanying data and metadata have not yet been deposited in a repository for long term preservation, although the team has selected to use the general-purpose Zenodo[33], at least until a more suitable platform is developed. Zenodo is not a disciplinary infrastructure and does not offer any specific function for 3D data or for cultural heritage data and metadata. However, this repository assigns a DOI to all deposited items, allows the inclusion of generic, high-level metadata (DataCite Metadata Schema, Dublin Core) and is otherwise well known and supported by the research community. We chose Zenodo because, in addition to be compliant with Open Science principles, it is already well known to all the partners of the project, is independent from the various institutions involved in the project while being available for all, and allowed us to create a community for the CHANGES-Spoke 4 project to group all the project outcomes under the same umbrella, instead of distributing them in a plethora of different repositories. Even if this is not an ideal solution, being Zenodo a repository which is not specific for Cultural Heritage data, in our context (i.e. Italy) there is not (yet) a disciplinary repository dedicated to Cultural Heritage research data. However, the project H2IOSC (https://www.h2iosc.cnr.it/) aims at creating a collaborative cluster of European distributed research infrastructures (CLARIN, DARIAH, E-RIHS, and OPERAS) involved in the humanities and cultural heritage sectors with operating nodes across Italy. Once H2IOSC provides a suitable repository for CH research data, we plan to move all the data and metadata produced in our process there.

[33] https://www.wikidata.org/

## 5. DISCUSSION AND CONCLUSION

### *5.1 FAIR-by-design digital twin: lessons from the use case*

In the previous section we have described in detail the process followed in digitising the temporary exhibition "Ulisse Aldrovandi and the Wonders of the World". This provides a good starting point to answer RQ1 ("What processes can be put in place to create a FAIR-by-design digital twin of a temporary exhibition?").

Before looking in more detail at how and why we consider our approach to be FAIR-by-design, it is worth stressing the importance of the preparatory work that went into conceiving and modelling the two tabular datasets (i.e. bibliographic data and process data), and especially the latter. Indeed, the results of this research project would be far less compliant with FAIR principles if the workflow had not been modelled in advance, and information regarding responsibilities (institutional and personal), important dates, tools and techniques used, had not been systematically and collaboratively collected and then converted and made available as Linked Open Data. This approach puts emphasis on research transparency, openness and accountability through the careful documentation of data collection and management techniques [47].

If we go back to the *FAIR Principles for Heritage Library, Archive and Museum Collections* we can look in turn at each of the three levels (Table 2) they have identified: the object level, the object metadata level and the metadata records level [1]. For each, we can assess how the Aldrovandi Digital Twin has met the relevant principles, if and how that has been done "by-design", and discuss the potential challenges encountered.

#### *5.1.1 Objects–in our case mostly 3D models*

**Findability: objects (physical and digital) have a globally unique persistent identifier and are described with metadata.**

We have described in the previous section the use of RDF to represent the metadata of the CHO, RAW, RAWp, DCHO and DCHOo, using standard disciplinary data models like CIDOC-CRM (see Section 4.2). In this context, permanent identifiers acting as IRIs–e.g., assigned by means of existing community-maintained services such as w3id.org, which is run by the W3C Permanent Identifier Community Group and provides a secure, permanent URL re-direction service for the Web applications–are assigned to the CHO, RAW, RAWp, DCHO and DCHOo. In addition, in our project, the DCHOo is assigned a unique alphanumeric identifier when uploaded into ATON (the web-based framework chosen for our project), and both DCHO and DCHOo will be assigned a DOI when uploaded into Zenodo.

**Table 2.** Summarising three levels in one table. Adapted from [1].

| | Objects | Metadata about the object (elementary level) | Metadata records |
|---|---|---|---|
| **Findable** | RDF is used as a data model to represent the metadata of the CHO, RAW, RAWp, DCHO and DCHOo, requiring the assignment of IRIs to each. Additionally, the description of these objects is facilitated through the use of CIDOC-CRM (via an application profile developed for this purpose). | Each CHO, RAW, RAWp, DCHO and DCHOo has its own IRI acting as permanent identifier (e.g., w3id.org) specified through metadata (see Objects Findability on the left). Additionally, each CHO, RAW, RAWp, DCHO and DCHOo may have other external persistent identifiers (PIDs).<br>Metadata also specify institutions responsible for holding the CHO. | Each metadata record is represented as a named graph and has its own IRI as a permanent identifier, specified through metadata. |
| **Accessible** | Ensured through version management, open and standard formats, and a streamlined plan to deposit objects in a trusted data repository (e.g., Zenodo), accompanied by clear and machine-readable metadata. | Through the use of a data repository (e.g. Zenodo), information about object availability and access options will be clearly stated in the metadata, in a machine-readable form. | Each metadata record is encoded as an RDF graph and made available in a triplestore equipped with a SPARQL endpoint for programmatic queries. |
| **Interoperable** | Open-source software is used whenever possible and objects are systematically exported into standard and open file formats (e.g., glTF, an interoperable open standard targeting interactive Web3D applications). | CIDOC-CRM and CRMdig are used as schemas to describe both physical and digital objects, and authority lists such as VIAF and Getty ULAN have been used as controlled terminology sources by referring to their terms through their own identifiers. | Metadata records include descriptive and contextual metadata of the CHO (titles, identifiers, descriptions) and the RAW, RAWp, DCHO and DCHOo (contextual information about their digitisation processes), ensuring a comprehensive representation in the context of cultural heritage acquisition and digitisation. |

**Table 2.** *Continued.*

| | Objects | Metadata about the object (elementary level) | Metadata records |
|---|---|---|---|
| **Reusable** | The acquisition step resulting in the RAW and the software activities that produced the RAWp, DCHO and DCHOo have all been assigned timestamps in the form of time intervals with a start date and an end date. The choice of licences of the objects is outside of our control but is implemented through the systematic inclusion of licensing information in a machine-readable form. | OPI has been rigorously recorded for the CHO, RAW, RAWp, DCHO and DCHOo. This includes detailed information of the institution holding the CHO and the people involved in the production of the RAW, RAWp, DCHO and DCHOo. | MRPI has been rigorously recorded for the metadata of the CHO, RAW, RAWp, DCHO and DCHOo. This includes detailed tracking of the agent responsible for data creation, the time of data creation, and the primary source of the data. This ensures accountability and provides a comprehensive historical record of the data, enhancing its reuse potential. |

Findability is designed into the system as the metadata of the CHOs were immediately collected, while the RAW, RAWp, DCHO and DCHOo were assigned relevant metadata, including an IRI, from the very beginning of the digitisation process. The choice of schema was carefully planned and, wherever possible, preceded the beginning of data collection.

**Accessibility: digital objects are permanently accessible by sustainable storage (hardware, storage medium), open universal access protocols, version management, backups.**

A vast range of technologies and procedures are used in the development and/or collection of 3D data. In our case study, along with the raw material (RAW) obtained by the acquisition step, we decided to keep three different derivative versions for each 3D model:

- Processed Raw Model (RAWp): the rough result of the photogrammetry or scanner software obtained by data processing excluding interpolation and geometry fixing;
- Digital Cultural Heritage Object (DCHO): the model obtained by interpolation and resolution of the geometry issues.
- Optimised Digital Cultural Heritage Object (DCHOo): the 3D model is optimised for real-time online interaction.

The RAWp and DCHO are stored in OBJ or FBX format, some of the most common formats for 3D models [29]. Each OBJ is exported together with the Material Texture Library (MTL) that defines the material specifics and the texture associated with the 3D model, stored in PNG or JPG format. Instead, the DCHOo is stored in glTF format and published on ATON.

RAWp, DCHO, and DCHOo will be made accessible through deposit on a trusted data repository–currently, in our project, most likely Zenodo [45]–that makes metadata openly available online and displays a clear, machine-readable licence for the data (see later under *Reusability*).

Accessibility by-design is ensured through version management, open and standard formats, and a streamlined plan to deposit objects in a trusted data repository.

**Interoperability: digital objects are stored in preferred or acceptable formats.**

This point has mostly been addressed in the paragraphs above. It is worth adding that glTF is an interoperable open standard targeting interactive Web3D applications that guarantees high interoperability with existing 3D platforms/services, re-use and integration of licensing information.

Every step of the workflow was preceded by a comprehensive review of the technologies currently in use in this fast evolving field. While open-source software was preferred, proprietary software was needed for specific tasks. To avoid being bound to any proprietary software application, we used as many standard formats as possible for all the types of research data produced (see Section 4.3).

**Reusability: digital objects have a date-timestamp and a licence for reuse, which is also available in a machine-readable form.**

The acquisition step and the software activities that produced the various digital objects have all been assigned timestamps in the form of time intervals with a start date and an end date (see Section 4.2).

Establishing the licence of each digitised object required a prompt and direct dialogue with the source institution, especially for the exhibition objects on loan, hence not directly held and managed by the University of Bologna. The licences currently in use for the majority of the digitised objects range from extremely permissive to more restrictive (e.g., Creative Commons Attribution Non-Commercial[34]) [45].

Reusability, while somehow limited by elements outside of our control, is implemented through the systematic inclusion of licensing information in a machine-readable form (see also next section).

#### *5.1.2 Metadata about the object (elementary level)*

**Findability: metadata specify the global persistent identifier (PID) of the object, which is used in all systems/databases that contain a description of the object in question, and metadata about specific objects are available via one or more searchable online repositories, catalogues, online databases, etc.**

[34] https://creativecommons.org/licences/by-nc/4.0/legalcode

As explained above, each CHO and DCHO is assigned from early on its own permanent identifier (an IRI) included into its own descriptive metadata record. Each CHO, RAW, RAWp, DCHO and DCHOo may have additional "external" persistent identifiers (PIDs), such as those assigned by the holding institution, or by a third-party (e.g., a DOI).

In our project, following deposit in a trusted data repository, each DCHO and DCHOo will also be assigned a DOI, which will be also specified in their metadata. Indeed, the deposit in an infrastructure of this kind adds a further layer of generic metadata that, while not adding particularly useful information for data re-use in research, ensures wider discoverability through online registries and search engines.

**Accessibility: metadata specify information about an object's availability, obtainability and/or access options.**

Through the use of a data repository, information about an object's availability and access options will be clearly stated in the metadata, in a machine-readable form (Creative Common licence, see above).

**Interoperability: metadata are available at least in one metadata schema appropriate for the specific type of object, are available in various additional generic standard data formats for other contexts, and contain references to other objects/authority files, by using other global persistent identifiers.**

This condition is certainly met – see Section 4.2 for a description of the management of object metadata as Linked Open Data (LOD). In our project, CIDOC-CRM and CRMdig have been used as main schemas to describe both physical and digital objects, and authority lists such as VIAF and Getty ULAN have been used as controlled terminology sources by referring to their terms through their own identifiers.

In addition to this disciplinary schema, appropriate for the specific type of object, the upload on Zenodo will add a further layer of generic metadata.

**Reusability: metadata specify the object's rights holder, OPI, and contain licence information referring to the object.**

Metadata specify the right holder of the CHO, RAW, RAWp, DCHO and DCHOo and contain licence information. As mentioned, in our project the licence of each RAW, RAWp, DCHO and DCHOo has been decided in agreement with the holding institution, while descriptive metadata have been made openly available (i.e. licenced using the CC0 waiver[35]).

OPI has been rigorously recorded for the CHO, RAW, RAWp, DCHO and DCHOo. This includes detailed information of the institution holding the CHO and the people involved in the production of the RAW, RAWp, DCHO and DCHOo. This ensures accountability and provides a comprehensive historical record of the RAW, RAWp, DCHO and DCHOo, in particular, enhancing its reuse potential (see Section 2.2).

[35] https://creativecommons.org/public-domain/cc0/

#### *5.1.3 Metadata records*

**Findability: metadata records have their own global persistent identifier.**

Each metadata record is represented as an RDF named graph and has its own IRI as a permanent identifier, created via w3id.org, specified through metadata.

**Accessibility: metadata records are machine readable and accessible using open universal protocols.**

Metadata are published as machine-actionable Linked Open Data. They are converted from an initial tabular format into a Resource Description Framework (RDF) format following specific community standards, i.e., CIDOC-CRM and related data models (see Section 4.2). Eventually, the Linked Open Data are made available in an RDF triplestore equipped with a SPARQL endpoint for programmatic queries.

**Interoperable: metadata records must be of sufficient quality.**

From the standpoint of an external researcher who wants to reuse the data (and therefore looks for as much information in the metadata as possible), the quality and completeness of the metadata records content is enough to describe the data and support its reuse. The metadata records include descriptive and contextual metadata of the CHO (titles, identifiers, descriptions, people and roles involved in their creation and management, techniques, types, etc.) and the RAW, RAWp, DCHO and DCHOo (contextual information about their digitisation processes, like dates, responsibilities, inputs and outputs, etc.), thus ensuring a focused yet comprehensive representation of the information a researcher seeks in the context of cultural heritage acquisition and digitisation (the CHO, its historical and cultural context, how it was digitised, when, how, by whom, and the RAW, RAWp, DCHO and DCHOo produced during the process).

**Reusability: metadata specify the metadata record's provenance (MRPI), the entity responsible for the metadata record, the licence for reuse (also available in a machine-readable form).**

MRPI has been rigorously recorded for both metadata describing the CHO, RAW, RAWp, DCHO and DCHOo. This includes detailed tracking of the agent responsible for data creation, the time of data creation, and the primary source of the data. This ensures accountability and provides a comprehensive historical record of the data, enhancing its reuse potential (see Section 4.2).

### *5.2 Main challenges encountered*

Our second research question investigates the main challenges in applying FAIR principles to 3D data in cultural heritage studies. Many of them were touched upon in the literature review, where we highlighted a number of gaps identified in previous studies.

It has been noted that metadata accompanying 3D models often do not allow a full assessment of the relevance of the model for further use scenarios [24]. Blundell et al. [25] call for community-driven work to develop and refine truly cross-disciplinary and widely interoperable metadata standards, together with further work on documenting and standardising the production of 3D models of cultural heritage in general.

Worryingly, a significant amount of digital assets is not made available to the scientific community but remains stored in private archives on personal computers, while the visibility of scientific platforms remains negligible compared to those that are commercially owned [24].

The multiplicity and complexity of 3D data certainly plays a role in their relative marginality in the cultural heritage sphere. Depending on the acquisition technique, different data types are obtained. In addition, 3D content generally goes through different phases and creators will have different versions of data or derivatives to store. This in turn entails:

- A multiplicity of file formats: there are currently more than 140, but this does not mean that all of them are or will remain popular [33]. Also, key features like scene details and animations are not compatible with all file formats, and these factors must be carefully considered when evaluating archiving formats.
- Data storage space, and related costs, reportedly one of the biggest challenges to date [29]. Indeed, preserving both raw and processed data is considered essential for the scientific reuse of 3D models, but can be extremely costly.

The choice of formats, as already explained, has been guided by a review of the most common and suitable formats, and a desire to stick to standard and non-proprietary formats.

We have described previously our internal policy regarding the conservation of three different versions for each 3D model (RAWp, DCHO and DCHOo) and the respective formats (glTF for DCHOo, OBJ or FBX for the others). At the time of this publication, we have not filled half of the storage space typically assigned to each file hosting cloud drive offered by the University of Bologna for a research project (33% out of 1 TB). Although not all project data have been uploaded yet, the selection of data and its format proved to be efficient. RAW files, which have been described as the most expensive data in 3D data preservation [29], occupy 46% of the storage space already in use, followed by the RAWp models obtained by processing operations (43%). The RAW and RAWp are therefore much more expensive than finished models: the DCHO files occupy 7%, while the DCHOo files do not even reach 1% of occupied space. The remaining 4% is occupied by project documentation, including photos and videos of the acquisition activity.

As previously mentioned, it is not always possible to stick to open-source software when working with 3D data (see Section 4.3). Converting data to standard formats is helpful but when converting raw material to non-proprietary formats, metadata embedded in the file can be lost and for this reason proprietary formats are often preferred [29]. In this context, choosing neutral formats that allow for structured and customised embedded metadata can facilitate the process (e.g. PLY). As of this publication, there is not a universally shared file format able to solve all the complexities involved, which makes interoperability even more challenging. See also Section 4.3 for a description of which formats and software tools were used in our case study.

Regarding 3D data publication, and despite the growing number of models available, there are still few online libraries and accessible download platforms. Additionally, commercial and non-commercial portals often fail to offer unique IDs or DOIs, provenance information or other pertinent metadata, links to additional resources, or even integrated 3D viewers (2D image or video preview is provided instead). These platforms should probably consider including the possibility to: create connections between models and historical records (as well as links for social media sharing and citation); create a DOI for each published 3D model; offer a customised and platform-integrated 3D model viewer; accurately track website traffic considering, if possible, offline and online usage [33].

For the publication of the 3D models of the Aldrovandi Digital Twin (i.e., the DCHOo files) we could rely on the open access web-based framework ATON, created by the CNR-ISPC in 2016 for the cultural heritage sector and built on open-source software and solid web standards [35].

Moving on to the long-term preservation of 3D cultural heritage data, we have described in Section 2.3.4 the results of a quick search on the Re3Data registry. It suggests that the availability of infrastructures for this purpose is patchy, perhaps because the long term archiving of cultural heritage 3D data is not yet a widespread practice in the arts and humanities research community. In our list of results, repositories devoted specifically to 3D data are exceedingly rare (three in total), as are institutional data repositories accepting 3D models (also three). In selecting a repository, we preferred a widely-used but general-purpose infrastructure for the deposit of all data and metadata (see Section 4.3). Zenodo offers the needed guarantees in terms of FAIRness and machine actionability, including displaying a clear reuse licence. It is possible that this choice might change in the future, should a discipline specific repository arise, for example in the context of the H2IOSC Project-Humanities and cultural Heritage Italian Open Science Cloud[36].

A very interesting point affecting Reusability–that we do not have the space to address in detail here–is whether it is possible to claim that the DCHO and DCHOo depicting real-world cultural heritage objects are not reproductions of these objects but rather new intellectual works in their own right. This discussion is not new to cultural heritage but remains extremely important as its results affect licensing decisions. Creating 3D models, in our view, requires craftsmanship and a high degree of interpretation, as it entails the reconstruction and juxtaposition of many different individual elements in order to build something that approximates reality but also adds something to it.

### 5.3 Peculiarities of 3D data and further research

We will conclude by summarising the new challenges that 3D models present for the cultural heritage sector and addressing the third and last research question: how are they different from traditional digital images from a data management perspective?

[36] https://www.h2iosc.cnr.it/

According to Moore et al. [29], although the form and requirements of the preserved items have changed, the addition of 3D data archiving does not affect the primary goals of digital repositories. Some of the data management challenges are indeed common to all disciplines and data types. They include choosing the most suitable file types, metadata standards, storage, licensing and rights, and making long-term curation decisions, taking into account technology potential and limits.

Although several cultural heritage institutions have been experimenting with 3D data creation and dissemination, it has not yet been widely embraced and supported in the GLAM sector. Among the museums that have made 3D data and multimedia available to the public we find the Smithsonian[37], the British Museum[38], the Naturhistorisches Museum Wien (NHMW)[39], the Museum fur Naturkunde Berlin[40]; among other types of institutions, it is worth mentioning the University of Amsterdam's 4D Research Lab[41].

Looking at 3D model scientific publishing specifically, the crucial role of metadata is gaining recognition but the lack of shared standards for the publication and exchange of digital heritage 3D assets limits the access and use of published resources. The multiplicity of 3D data does not facilitate standardisation, which may indeed prove even longer and more complex to achieve than for 2D data. Furthermore, due to the different 3D data creation methods and environments of usage, there is no shared file format for 3D model preservation and interoperability between different software is a long way to go. To exploit 3D data potential for the analysis of cultural heritage objects, efforts are needed in terms of technology and process standardisation to make data sustainably accessible.

The size of 3D content is one of the biggest challenges concerning the long-term preservation of raw material (see Section 5.2). Finally, the functions offered both by public and commercial 3D platforms for 3D model retrieving, access, and analysis are still limited and under development (see Section 2.3.3) while very few data repositories for long-term archiving seem to be available (see Section 2.3.4). In this context, research-based on 3D digital heritage projects still lacks critical insights and cooperative decision-making processes to design data curation in a FAIR-compliant way.

A first step in this direction could be the creation of a FAIR Implementation Profile (FIP) within the cultural heritage community working with 3D data. FIPs register implementation choices made to adopt FAIR principles and can accelerate the adoption of community-driven and community-specific consensus around FAIR by-design data production and management practices [48].

[37] https://3d.si.edu/explore?edan_local=&edan_q=&edan_fq%5B0%5D=media_usage%3ACC0 and https://sketchfab.com/

[38] https://sketchfab.com/britishmuseum/models and https://www.bmimages.com/3d-scans.asp

[39] https://www.nhm-wien.ac.at/en/museum_online/3d. Interestingly, the institution also has a Data Repository run by the museum's own publishing arm in collaboration with the NHMW Central Research Laboratories and IT department: http://datarepository.nhm-wien.ac.at/. This archive contains datasets, images and textual sources, but not the aforementioned 3D models

[40] https://portal.museumfuernaturkunde.berlin/ and https://sketchfab.com/VisLab

[41] https://4dresearchlab.nl/

## ACKNOWLEDGMENTS

This work has been partially funded by Project PE 0000020 CHANGES-CUP B53C22003780006, NRP Mission 4 Component 2 Investment 1.3, Funded by the European Union-NextGenerationEU.

## AUTHOR CONTRIBUTIONS

**Sebastian Barzaghi:** Data curation, Methodology, Writing–original draft, Writing–review & editing.

**Alice Bordignon:** Investigation, Methodology, Visualization, Writing–original draft, Writing–review & editing.

**Bianca Gualandi:** Conceptualization, Writing–original draft, Writing–review & editing.

**Ivan Heibi:** Methodology, Writing–original draft, Writing–review & editing.

**Arcangelo Massari:** Methodology, Software, Writing–original draft, Writing–review & editing.

**Arianna Moretti:** Methodology, Software, Writing–original draft, Writing–review & editing.

**Silvio Peroni:** Investigation, Methodology, Supervision, Writing–review & editing.

**Giulia Renda:** Software, Writing–original draft, Writing–review & editing.

## AUTHOR BIOGRAPHY

**Sebastian Barzaghi** is research fellow at the Department of Cultural Heritage of the University of Bologna. Graduated in 2020 from the Digital Humanities and Digital Knowledge (DHDK) master's program at the University of Bologna, he currently works at the FrameLAB research laboratory on various projects related to the application of Linked Open Data methodologies and technologies. Specifically, his research interests focus on how Semantic Data Modeling and Open Science can improve scholarly endeavours (research, teaching and communication) in the Humanities and Cultural Heritage domains.
E-mail: sebastian.barzaghi2@unibo.it; ORCID: 0000-0002-0799-1527

**Alice Bordignon** is a research fellow at the Department of Classical Philology and Italian Studies (FICLIT) at the University of Bologna. She holds a Bachelor's degree in Art History and a Master's degree in Digital Humanities. She is currently involved in the PNRR CHANGES / Spoke 4 (Virtual Technologies for Museum and Art Collections) project, where she conducts activities including reality-based 3D acquisition, processing, rendering, and web optimisation of cultural heritage using photogrammetry, 3D modelling, and Web3D technologies.
E-mail: alice.bordignon2@unibo.it; ORCID: 0009-0008-3556-0493

**Bianca Gualandi** is data steward and PhD candidate at the University of Bologna. Her research focuses on what it means to open up research in the Humanities, and on the concept of 'data' in this field. Her daily job consists in advocating for and facilitating FAIR data management practices across the entire research lifecycle, especially within the Cultural Heritage sector.
E-mail: bianca.gualandi4@unibo.it; ORCID: 0000-0001-8202-8493

**Ivan Heibi** is an assistant professor at the Department of Classical Philology and Italian Studies (FICLIT) of the University of Bologna. His work mainly concerns the use of Semantic Web technologies in the analysis of bibliographic entities and the behaviours/dynamics of the scholarly publishing. He is a member of the Research Centre for Open Scholarly Metadata (https://openscholarlymetadata.org/) and The Digital Humanities Advanced Research Centre (DHarc, https://centri.unibo.it/dharc).
E-mail: ivan.heibi2@unibo.it; ORCID: 0000-0001-5366-5194

**Arcangelo Massari** holds a Bachelor's degree in Humanities and a Master's degree in Digital Humanities and Digital Knowledge from the University of Bologna. He also works as a developer at OpenCitations, a not-for-profit organization promoting open scholarship. During his tenure as a research fellow, Arcangelo contributed significantly to the development of OpenCitations Meta, a comprehensive database that stores and provides access to bibliographic metadata for publications indexed by OpenCitations. Currently, Arcangelo is pursuing a PhD in Cultural Heritage in the Digital Ecosystem at the University of Bologna, while simultaneously working towards a joint PhD in Engineering Technology at the University of Leuven. He is an active participant in the Doctoral Programme in Cultural Heritage, coordinated by UnaEuropa. His doctoral research centers on the development of tailored interfaces designed to streamline the curatorial workflows of archivists, librarians, and museologists. This work aims to create versatile tools capable of supporting a wide variety of data formats and database systems used in the preservation of cultural heritage.
E-mail: arcangelo.massari@unibo.it; ORCID: 0000-0002-8420-0696

**Arianna Moretti** is a Ph.D. student in the 'Cultural Heritage in the Digital Ecosystem' program at the Department of Classical Philology and Italian Studies of the University of Bologna and in the 'Una Her Doc' European doctoral program in Cultural Heritage by Una Europa. Her primary research focus lies in the automation of metadata crosswalks between different data models and file formats. Specifically, she is dedicated to creating a highly reusable workflow for exchanging variably modeled metadata in the cultural heritage domain, with a particular emphasis on RDF serialization formats. This research interest extends to her involvement in Spoke 4 of the PNRR Extended Partnership 5 (Cultural Heritage Active Innovation for Sustainable Society-CHANGES). Additionally, she collaborates with OpenCitations under the Research Centre for Open Scholarly Metadata at the University of Bologna, where her responsibilities include performing crosswalks from the data source data models to the OpenCitations data model.
E-mail: arianna.moretti4@unibo.it; ORCID: 0000-0001-5486-7070

**Silvio Peroni** holds a Ph.D. in Computer Science and is an Associate Professor at the Department of Classical Philology and Italian Studies, University of Bologna. He is director of OpenCitations, director of the Research Centre for Open Scholarly Metadata, and director of the Second Cycle Degree in Digital Humanities and Digital Knowledge (DHDK). His current works concern theoretical studies and technical implementation of tools to foster semantic interoperability of Open Science services and infrastructures, the empirical analysis of the nature of scholarly citations, bibliometrics and scientometrics studies, visualisation and browsing interfaces for semantic data, and the development of ontologies to manage, integrate and query bibliographic information and cultural heritage data.
E-mail: silvio.peroni@unibo.it; ORCID: 0000-0003-0530-4305

**Giulia Renda** is a PhD student in Cultural Heritage in the Digital Ecosystem at the University of Bologna. She holds a Bachelor's degree in History and a Master's degree in Digital Humanities. Her research focuses on integrating Semantic Web technologies to enhance access to and interaction with digital cultural heritage. She specialises in developing intuitive software solutions for cultural data representation and visualisation. During her time with Polifonia H2020, she contributed to MELODY, a tool for creating data stories with Linked Open Data, and the Polifonia Web Portal, a platform for aligning data and enabling discovery in music heritage.
E-mail: giulia.renda3@unibo.it; ORCID: 0000-0003-2990-0021